\documentclass[aps,showpacs,floatfix,twocolumn]{revtex4-1}
\usepackage{graphicx}
\newcommand{\be}{\begin{eqnarray}}
\newcommand{\ee}{\end{eqnarray}}
\def\beq{\begin{equation}}
\def\eeq{\end{equation}}

\begin{document}
\title{Localization in fractal and multifractal media}
\author{Antonio M. Garc\'{\i}a-Garc\'{\i}a}
\affiliation{CFIF, Instituto Superior T\'ecnico, 
Universidade T\'ecnica de Lisboa, Av. Rovisco Pais, 1049-001 Lisboa, Portugal}
\author{Emilio Cuevas}
\affiliation{Departmento de F\'{\i}sica, Universidad de Murcia,
E-30071 Murcia, Spain}
\begin{abstract}
The propagation of waves in highly inhomogeneous media
is a problem of interest in multiple fields including seismology, acoustics and electromagnetism.
It is also relevant for technological applications such as the design of sound absorbing materials or the fabrication of optically devices for multi-wavelength operation. A paradigmatic example of a highly inhomogeneous media is one in which the density or stiffness has fractal or multifractal properties. We investigate wave propagation in one dimensional media with these features. We have found that, for weak disorder, localization effects do not arrest wave propagation provided that the box fractal dimension $D$ of the density profile is $D \leq 3/2$. This result holds for both fractal and multifractal media providing thus a simple universal characterization for the existence of localization in these systems. Moreover we show that our model verifies the scaling theory of localization and discuss practical applications of our results.
 \end{abstract}
\pacs{05.45.Df,73.20.Fz, 05.60.Cd,46.40.Cd,46.65.+g}
\maketitle
Wave propagation \cite{locwaves,anderson} in disordered or highly inhomogeneous media is a recurrent topic of research in physics. This is not surprising due to its broad range of potential applications.
The propagation of electromagnetic waves in highly inhomogeneous fractal media  \cite{elec1,elec2,elec3,elec4} was studied experimentally in \cite{elec2,elec3}. Anderson localization effects in acoustic waves have been observed in experiments \cite{acousrev} and numerical simulations \cite{acous2}. Seismic waves \cite{iran} also propagate in a highly inhomogeneous medium as there is a growing consensus \cite{seisrev,seis1} that the distribution of fractures and densities in the
earth inner structure has a fractal distribution.
A detailed understanding of this problem is also relevant for the fabrication of optically active devices for multiwavelength operation \cite{elec1}, the fabrication of a multiple gap and multiple pass band micro-strip resonator filter \cite{elec4}, the design of sound absorbing materials and the nondestructive characterization of fractured materials \cite{acous1}.

 Rigorous theoretical results are known only for one-dimensional (1d) matter waves in correlated random potentials $V(n)$.
Kotani \cite{kotani} demonstrated that Anderson localization -- stop of diffusion and exponential decay of eigenfunctions -- occurs for any energy and amount of disorder \cite{anderson,locwaves} provided that correlations decay as $\langle
V(n)V(0) \rangle \propto 1/n^\alpha$ with $\alpha > 0$ or faster. Anderson localization effects are also suppressed as the degree of differentiability of the potential increases \cite{us}. For other recent studies of localization in correlated 1d potentials we refer to \cite{physcorr}.
It is unclear to what extent these results still hold in the case of classical waves in a fractal-like media. \\
The main aim of this paper is to answer this question. We investigate what are the most general circumstances in which a perturbation in a highly heterogeneous 1d medium, described by a fractal or multifractal density, propagates without ever experiencing Anderson localization.\\
Our main result is that a perturbation will propagate indefinitely in 1d provided that the box fractal dimension of the medium density/stiffness fluctuations is $D \leq 3/2$.  These results are still valid if the density profile is multifractal.\\
\begin{figure}[!ht]
\includegraphics[width=0.78\columnwidth,clip]{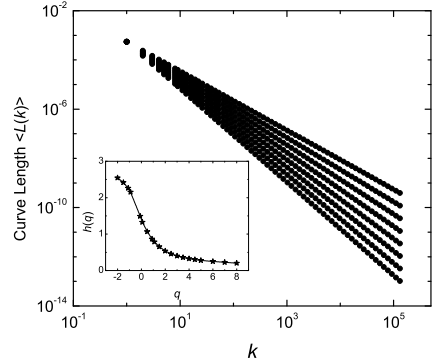}
\caption{Average length $\langle L(k) \rangle$ of the medium over an interval of width $k$. By definition the slope of the curve is the box dimension D. The results correspond to a density profile given by a fBm
with Hurst exponent H. H ranges from $0.1$ to $0.8$, in steps of $0.1$, from botton to top. In all cases $D = 2 - H$. Inset: Generalized Hurst exponent $h(q)$ ($h(2) \equiv H$) \cite{stanley} that describes the scaling of a multifractal density profile generated by a multiplicative random cascade process (see text).}
\label{fig1}
\end{figure}
{\it Monofractal density.-}
We investigate
a discretized version of the 1d scalar wave equation,
\be
\label{waveq}
\rho_n(\psi_{n+1} -\psi_{n}) - \rho_{n-1}(\psi_{n} -\psi_{n-1}) +\omega^2 \psi_n = 0,
\ee
 where $\omega$ are the eigenfrequencies of the system, $\rho_n = \rho_0 ( 1 + \eta_n)$ the ratio between the stiffness and the density of the material and $\rho_0$ is the average value. The fluctuations $\eta_n$ are described by a fractal function with box fractal dimension $D$. For the sake of simplicity we study localization properties in a narrow region of frequencies 
 outside the origin $\omega \approx 0$ as a function of the box dimension D and the system size L. 
We will restrict ourselves to the region of weak disorder. This a natural choice as our main motivation is to put forward a characterization for the existence of delocalized eigenmodes.     
As examples of density profiles with a monofractal spatial --labelled by $n$-- distribution we employ: (i) the Weierstrass curve \be
\label{wei}
\eta_n^{we} = \sum_{k}^{\infty} \cos(\gamma^k n /L +\phi_k)/\gamma^{(2-D)k},\ee with $\gamma > 1$, $\phi_k $ is a box distributed random number $\in [0, 2\pi]$ and $n$ is a spatial variable. The box fractal dimension is $D^{we} = D$ \cite{falconer, cascade};  
(ii) a fractional Brownian motion (fBm). This is a generalization of the standard Brownian motion characterized by stationary increments -- though the process itself is non-stationary--, self-similarity, and a variance $\sigma^2 \propto n^{2H}$ where $H$ is the Hurst exponent. We note that, due to its self-similar character,  every realization 
of the process is a fractal curve with box-dimension $D = 2 - H$ \cite{falconer}. 
In both cases in order to have a well defined continuous limit we rescale the potential such that the average is $<\eta> = 0$. The standard deviation $\sigma_\eta = \sqrt{< \eta^2 >}$ controls the strength of disorder. The box-dimension $D$ for a given
$H$ is obtained by the method presented in Ref. \cite{higuchi}. It basically consists in
calculating the average curve length $\langle L(k) \rangle$ over an interval of width $k$. If
$ \langle L(k) \rangle \propto k^{-D}$, then the curve is fractal with a box-dimension
$D$. In Fig. \ref{fig1} we show that for the fBm $D_{\rm fBm} = 2 - H$.  
\begin{figure}[ht!]
\includegraphics[width=0.78\columnwidth,clip]{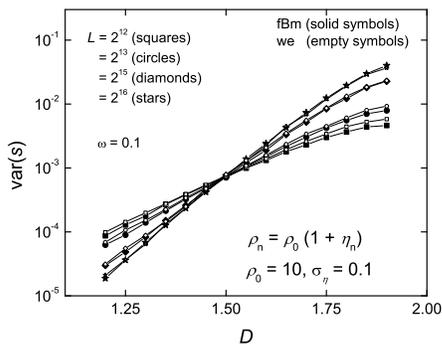}
\caption{(color online)  ${\rm var(s)}$, as a function of $D$ for $\omega \approx 0.1$, $\sigma_{\eta} = 0.1$ and different sizes $L$ obtained by numerical diagonalization of the wave equation Eq.(\ref{waveq}) for a density profile described by the Weiertrass function Eq.(\ref{wei}), and a fractional Brownian motion (see text for details). 
In both cases ${\rm var(s)}$ increases with the system size -- a signature of Anderson localization -- only for $D > 3/2$. This indicates that $D = D_c = 3/2$ is the maximum box dimension for which a band of extended modes can occur.}
\label{fig2}
\end{figure}
We compute the eigenmodes $\omega$ of Eq.(\ref{waveq}) by using standard numerical diagonalization techniques. For a given disorder $\sigma_{\eta}$ and frequency window the number of eigenmodes
obtained is at least $2 \times 10^5$. In order to determine the importance of Anderson localization effects
we carry out a finite size scaling analysis of the spectrum \cite{sko}.
The variance $\rm var(s)$ of the eigenmode spacing distribution $P(s)$ is chosen as the scaling variable \cite{sko}.
$P(s)$ is the probability of finding two neighbouring
eigenmodes at a distance $s = (\omega_{i+1} - \omega_{i})/\Delta$ and 
\be
{\rm var(s)} \equiv \langle s^2 \rangle -
\langle s \rangle^2
  = \int_0^{\infty}ds ~s^2 P(s)- 1, \ee
where $\langle \dots \rangle$ denotes frequency and ensemble averaging and $\Delta$ is the local mean eigenmode spacing.
In case that propagation is not stopped by localization effects,
${\rm var(s)} \approx 0.273$ for diffusive motion and ${\rm var(s) = 0}$ for ballistic motion.
By contrast ${\rm var(s)} = 1$ indicates that a perturbation cannot propagate indefinitely in the medium.
A value of ${\rm var(s)}$ that increases (decreases) with
 system size signals that the perturbation will eventually (never) get localized \cite{sko}.\\
\begin{figure}[!ht]
\includegraphics[width=0.78\columnwidth,clip]{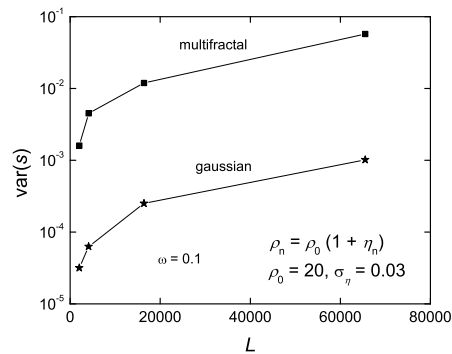}
\caption{${\rm var(s)}$, as a function of the system size $L$, fixed disorder $\sigma_{\eta} = 0.03$ and $\omega \approx 0.1$ for a density profile described by a Gaussian white noise and a multifractal one generated by a multiplicative random cascade process (see text). In both cases the box dimension is $D = 2$ and the two point correlation function $\langle \eta_n \eta_0 \rangle \propto \delta(n)$. Since $D > 3/2$ all modes are localized. However localization effects are stronger (${\rm var(s)}$ is larger) for the medium whose density is multifractal.}
\label{fig3}
\end{figure}
The results for $\rm var(s)$ for the fBm and the Weiertrass density profile, at fixed $\omega$ and disorder are shown in Fig. \ref{fig2} as a function of $D$.
 For $D > 3/2$ the variance $\rm var(s)$ increases with the system size. This indicates stop of wave propagation. By contrast
 for $D < 3/2$, $\rm var(s)$ decreases with the system size thus suggesting that, in this case, localization never occurs.
Therefore in fractal media the transport properties for sufficiently weak disorder are to a great extent controlled by
the fractal box dimension $D$ of the density/stiffness.
A natural question to ask is the degree of universality of this characterization in terms of $D$. In order to answer this question we investigate density profiles with multifractal properties. In this case the box dimension can still be defined but the scaling properties of the density profile depend on an infinite number of multifractal dimensions \cite{falconer}. It is thus unclear whether other fractal dimensions control localization related effects.\\
{\it Multifractal density generated by a cascade process.-}
We first study the effect of multifractality in a density profile whose two point correlation is $\langle \eta_n \eta_0 \rangle \propto \delta(n)$. 
Our motivation is to estimate the quantitative effect of a multifractal media by comparing it with one, an uncorrelated Gaussian noise,  which is not multifractal but have the same two point correlation function.
The multifractal density fluctuations are generated iteratively by a variant of a multiplicative random cascade process \cite{cascade}. The length of the process doubles in each iteration. Initially the density consists just of one value $\eta^{(0)} = \eta_0$.  In the next generation it has two values, $\eta^{(1)} = m_1^{(1)} \eta_0$ and $\eta^{(2)}= m_2^{(1)} \eta_0$ where $m_i$ are Gaussian random numbers of
zero mean and unit variance. After $k$ iterations the density will have $2^k$ different values given by, $\eta^{(k)}_{2l-1}=\eta_l^{(k-1)}m_{2l-1}^{(k)}$ and $\eta_{2l}^{(k)}=\eta_{l}^{(k-1)}m_{2l}^{(k)}$. Following \cite{stanley} we investigate the scaling properties of these data by using the multifractal detrended fluctuation analysis.
The resulting density is characterized by a generalized Hurst exponent $h(q)$ (see \cite{stanley} for a definition and Fig. 1).
For a monofractal curve, $h(q) = H$ is independent
of $q$. For this multifractal distribution, the box dimension is $D = 2$ as that of a Gaussian uncorrelated disorder. According to our previous characterization, localization should occur for any frequency and amount of disorder.
A finite size scaling analysis (not shown), as the one performed in the monofractal case, confirms this prediction. 
This result might suggest that localization effects in both media are similar. However this is not the case. Multifractal corrections, which arises in higher order correlation functions, play an important role. 
In order to illustrate this we compare in Fig. \ref{fig3} ${\rm var(s)}$ for both models as a function of $L$ at fixed $\omega$ and $\sigma_{\eta}$. We observe that: a) in both models ${\rm var(s)}$ increases with the system size. This is a signature of localized modes; b) Multifractal scaling enhances localization effects since ${\rm var(s)}$ is two order of magnitude larger in the this case. It is unclear to what extent this behaviour is universal. Multifractality leads to a broad range of scaling patterns. Therefore we cannot discard that in certain cases a suppression of localization effects can be observed. From our point of view the main conclusion of this analysis is that, in order to attain a good understanding of the transport properties in a highly inhomogeneous medium, is very important to know in detail the full distribution function that describes the medium density and not only its first moments.\\ 
{\it Multifractal density generated by a generalized random walk.-}
We now move to the case of a multifractal density in which $h(2)$ can be tuned such that the box dimension $D$ changes. Our motivation here is to test whether, as in the monofractal case, absence of localization is only observed for $D < 3/2$. 
For this purpose we study the multifractal random walk introduced in \cite{mrw}. The strength of disorder is set as in the monofractal case. Here we just state the definition of the model and refer to \cite{mrw} for details,
\be
\eta_n^{mul} = \sum_{i=1}^{n} \eta_i^{id}e^{r(i)},
\ee
where $\eta^{id}$ is a fractional Gaussian
noise,  $r(i)$ is a Gaussian correlated noise with $\langle r^2 \rangle = \lambda^2  \ln (L_0)$ and $\langle r(i)r(j)\rangle = \lambda^2 \ln f(|i-j|)$ with $\lambda$ a free parameter describing the strength of the multifractal scaling, $L_0$ is the largest scale for which multifractal scaling is observed and $f(|i-j|)= \frac{L_0}{1+|i-j|}$ for $|i-j| \leq L_0 -1$ and the unity otherwise.
With these definitions the resulting fluctuations of the density profile $\eta^{mul}$ are multifractal with a set of multifractal dimensions that depends on $\lambda$. As in the monofractal case we have carried out a finite size scaling analysis of the spectrum.
The results for $\rm var(s)$ (see Fig. \ref{fig4}), indicate that propagation is arrested only for $D > 3/2$ since in this region $\rm var(s)$ decreases with the system size. Therefore our characterization for the existence of extended modes is still valid for media with a multifractal density.
\begin{figure}[!ht]
\includegraphics[width=0.78\columnwidth,clip]{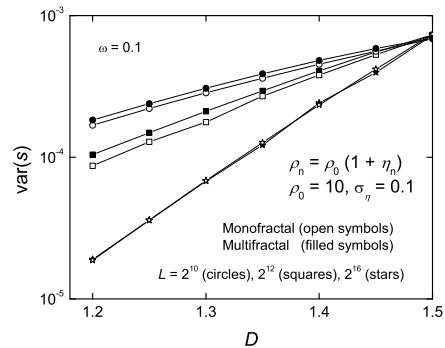}
\caption{${\rm var(s)}$, as a function of $D < 3/2$ for $\omega \approx 0.1$, $\sigma_\eta = 0.1$ and different $L$ for a density profile described by a multifractal random walk with $\lambda^2 = 0.1$, $L_0 = 2^{10}$ and a monofractal with the same box dimension $D$. Multifractality enhaces localization effects but the critical box dimension is still $D = D_c = 3/2$.}
\label{fig4}
\end{figure} 
This together with the previous results on monofractal media strongly suggests that the value of the box dimension $D$ provides a  characterization for the absence of Anderson localization in 1d media.\\
Finally we investigate whether our findings are consistent with the one parameter scaling theory \cite{one}.  According to this theory $\rm var(s)$ can be expressed by
a one-parameter scaling function
${\rm var(s)} =f(L/\xi (D))$
where the scaling parameter $\xi$ is the localization length
in the localized regime, and the correlation length in the
extended regime. This relation implies that in a log-log plot of $\rm var(s)$ 
versus $L$ all data should collapse in a common curve when translated by
an amount $\ln \xi(D)$ along the horizontal axis.
This curve has a single branch when there is no transition, while it
develops two separate branches when a transition is present.
In Fig. (\ref{fig5}) we plot $\rm var(s)$ as a function of $L$ for different values of $D$ at a fixed $\omega$.
The obtained result is fully consistent with the scaling hypothesis for $\rm var(s)$.
 Therefore the one-parameter scaling theory is still valid for this type of systems. \\   
\begin{figure}[!ht]
\includegraphics[width=0.78\columnwidth,clip]{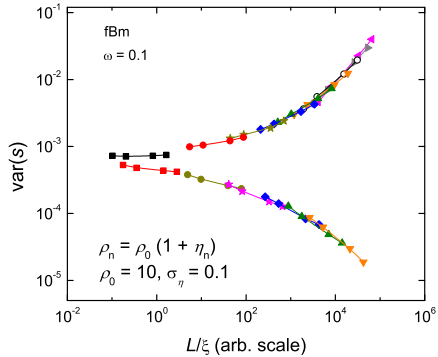}
\caption{(color online) 
$\rm var(s)$ a function of the system size for $\omega \approx 0.1$ and $\sigma_\eta = 0.1$ for a density described by a fBm.
Each colour/symbol corresponds to a different $D$. The observed data collapse is fully consistent with the one-parameter scaling assumption for
$\rm var(s)$ \cite{one} (see text). 
 }
\label{fig5}
\end{figure}
{\it Applications.-}
One of the main motivations to study wave transport in fractal and multifractal media is its relevance in many realistic situations. We now address some of these applications.
The distribution of fractures and densities in the
earth inner structure is believed to be fractal/multifractal \cite{seisrev,seis1}.
 Therefore a more comprehensive understanding of the transport properties in such a medium
 would not only help characterize the earth inner structure but also would be relevant for practical problems such as
 the minimization of earthquakes effects and the detection of possible
large-scale reservoir of oil and  gas \cite{seisapp}. We note that our results provide a characterization of the conditions that a fractal medium must meet in order that a wave can propagate through it.\\
A detailed understanding of the propagation of electromagnetic waves in a highly inhomogeneous medium is a key ingredient in the fabrication of different devices \cite{elec1,elec4} (see introduction). In this case our characterization would be useful to optimize its capabilities.
Likewise a comprehensive theory of the attenuation of acoustic waves in a fractal porous medium would improve dramatically the design of sound absorbing materials. \\
The study of matter waves in cold atoms settings could also benefit from these findings.
In this case the system is a tight-binding
Anderson model with a fractal-like potential. This type of potentials might be realized experimentally by using a holographic
mask combined with spatial
light modulators \cite{mask}. \\
To conclude, we have studied wave propagation in a 1d fractal media.
Our main results are: a) the fractal dimension of the potential $D$ controls the strength of localization effects;
b) only for $D < 3/2$ wave propagation is not arrested by localization effects; c) multifractal corrections do not modify this result but have a profound impact in the transport properties, d) scaling theory still applies in these systems.

We acknowledges financial support from  FEDER
and the Spanish DGI through Project No.
FIS2007-62238.


\vspace{-5mm}
\begin{thebibliography}{99}
\vspace{-5mm}
\bibitem{locwaves}F. Delyon, H. Kunz and B. Souillard, J. Phys. A {\bf 16}, 25 (1983);P. W. Anderson, Phil. Mag. B {\bf 52}, 505 (1985);S. John, Phys. Rev. Lett. {\bf 58}, 2486 (1987).
\bibitem{anderson}P. W. Anderson, Phys. Rev. {\bf 109}, 1492 (1958).
\bibitem{elec1}M. Hiltunen et al., J. Light. Technol., {\bf 25}, 1841, (2007).
\bibitem{elec2}W. J. Wen et al, Phys. Rev. lett. {\bf 89}, 223901 (2002).
\bibitem{elec3} M. W. Takeda et al., Phys, Rev. Lett. {\bf 92}, 093902 (2004).
\bibitem{elec4}S. Wang et al., Appl. Phys. {\bf A85}, 159, (2006).
\bibitem{acousrev}S. He and J.D. Maynard, Phys. Rev. Lett., {\bf 57}, 3171 (1986);J. D. Maynard, Rev. Mod. Phys., {\bf 73}, 401 (2001).
\bibitem{acous2}A. Esmailpour et al., Phy. Rev. B, {\bf 78}, 134206 (2008).
\bibitem{iran}A. Bahraminasab et al., Phys. Rev. B {\bf 75}, 064301 (2007);S. A. Shapiro, Geophys. J. Int. {\bf 110}, 591 (1992);S. Crampin, J. Pet. Sci. Eng. {\bf 24}, 29 (1999); A. Feustel, Tectonophysics {\bf 289}, 31 (1998);P. Manshour, Phys. Rev. Lett. {\bf 102} 014101 (2009).
\bibitem{seisrev}S. Lovejoy et al., Nonlin. Proc. Geophys., {\bf 14}, 465, (2007).
\bibitem{seis1}M. Pilkington and J. P. Todoeschuck, Geophys. Res. Lett. {\bf 31}, L09606 (2004).
\bibitem{acous1}S. P. Pride et al., Phys. Rev. Lett. {\bf 97}, 184301 (2006).
\bibitem{kotani}S. Kotani, {\it Stochastic Analysis}, Editor K. Ito, North-Holland, Amsterdam (1984), pp. 225–247;S. Kotani et al., Commun. Math. Phys. {\bf 112}, 103 (1987).
\bibitem{us} A. M. Garc\'{\i}a-Garc\'{\i}a and E. Cuevas, Phys. Rev. {\bf B 79}, 073104 (2009).
\bibitem{physcorr}T. Kaya,
Eur. Phys. J. B {\bf 55}, 49 (2007);E. Gurevich and O. Kenneth 
Phys. Rev. A {\bf 79}, 063617 (2009); A. Iomin 
Phys. Rev. E {\bf 79}, 062102 (2009); A. Sheikhan et al., 
Phys. Rev. B {\bf 80}, 035130 (2009); C. Monthus and T. Garel
Phys. Rev. E {\bf 81}, 011138 (2010).  
\bibitem{one}E. Abrahams et al., Phys. Rev. Lett. {\bf 42}, 673 (1979).
\bibitem{higuchi} T. Higuchi, Physica D {\bf 31}, 277 (1988).
\bibitem{grun}R. Gorenflo et al., Nonlinear Dynamics {\bf 29}, 129 (2002);
H. E. Roman et al., Phys. Rev. E. {\bf 78}, 031127 (2008).
\bibitem{sko}B. I. Shklovskii et al., Phys. Rev. B {\bf 47}, 11487 (1993);E. Cuevas, Phys. Rev. Lett. {\bf 83}, 140 (1999).
             
\bibitem{falconer}K. Falconer, {\textit{'Fractal Geometry'}}, Chichester, J. Wiley and Sons (1990); P. Flandrin et al., IEEE Trans. Inform. Theory, {\bf 38}, 910, (1992). 
\bibitem{cascade}J. Feder, {\textit{'Fractals'}} (Plenum, New York, 1988);M. Greiner et al., Phys. Rev. Lett. {\bf 80},
5333 (1998); M. I. Bogachev et al., Phys. Rev. Lett. {\bf 99}, 240601 (2007).
\bibitem{stanley}J. W. Kantelhardt et al., Physica {\bf A316}, 87 (2002).
\bibitem{mrw}J. F. Buzy et al., Phys. Rev. E {\bf 66}, 056121 (2002);E. Bacry et al., Phys. Rev. E {\bf 64}, 026103 (2001).
\bibitem{seisapp}S. Crampin, J. Pet. Sci. Eng, {\bf 24}, 29 (1999); A. Feustel, Tectonophysics, {\bf 289}, 31 (1998).
\bibitem{mask}M. Mutzel et al., Phys. Rev. Lett. 88, 083601 (2002).



\end{thebibliography}
\end{document}